\documentclass{nature} 
\usepackage{amsmath,amssymb} 
\usepackage{graphicx}
\usepackage{epstopdf} 
\usepackage{caption} 
\usepackage{color}
\newcommand{\bk}{\textbf{k}} \newcommand{\bq}{\textbf{q}}

\usepackage{color}
\definecolor{sb}{RGB}{0,100,180}
\definecolor{replace}{RGB}{120,120,120}

\begin{document}
\bibliographystyle{naturemag} 

\title{Quantifying many-body effects by high-resolution Fourier transform scanning tunneling spectroscopy}

\author{S. Grothe,$^{1, 2,*}$ S. Johnston,$^{1, 2,*}$ Shun Chi,$^{1, 2}$ P.
Dosanjh,$^{1, 2}$ S. A. Burke,$^{1,2,3}$ and Y. Pennec$^{1,2}$} 

\maketitle

\begin{affiliations} 
\item{Department of Physics and Astronomy, University of British Columbia, Vancouver BC, Canada V6T 1Z1} 
\item{Quantum Matter Institute, University of British Columbia, Vancouver BC, Canada V6T 1Z4} 
\item{Department of Chemistry, University of British Columbia, Vancouver BC, Canada V6T 1Z1}
\end{affiliations}

\date{\today}

\begin{abstract} Many-body phenomena are ubiquitous in solids, as 
electrons interact with one another and the many excitations arising
from lattice, magnetic, and electronic degrees of freedom. These
interactions can subtly influence the electronic properties of materials ranging
from metals,\cite{Migdal} exotic materials such as graphene, 
\cite{Bostwick2006,Bostwick2010} 
and topological insulators,\cite{ZhuPRL2012} 
or they can induce new phases of matter, as in conventional\cite{McMillan} and unconventional
superconductors,\cite{Lanzara,Dahm,Byczuk,Graf} 
heavy fermion systems,\cite{FiskNature1986} 
and other systems of correlated
electrons. As no single
theoretical approach describes all such phenomena, the development of versatile
methods for measuring many-body effects is key for understanding these
systems. To date, angle-resolved photoemission spectroscopy (ARPES) has been
the method of choice for accessing this physics by directly imaging 
momentum resolved electronic structure.\cite{Bostwick2006,Bostwick2010,
ZhuPRL2012,Lanzara,Dahm,Graf}
Scanning tunneling microscopy/spectroscopy (STM/S), renown
for its real-space atomic resolution capability, can also 
access the electronic structure in momentum space  
using Fourier transform
scanning tunneling spectroscopy (FT-STS).\cite{SprungerScience1997,
Petersen, McElroyNature2003, HoffmanScience2002} 
Here, we report a high-resolution
FT-STS measurement of the Ag(111) surface state, revealing fine structure in  
the otherwise parabolic electronic dispersion. This deviation is induced 
by interactions with lattice vibrations 
and has not been previously resolved by any technique. This study advances 
STM/STS as a method for quantitatively probing many-body
interactions. Combined with the spatial sensitivity of STM/STS, this technique
opens a new avenue for studying such interactions at the nano-scale.
\end{abstract}

Non-interacting electrons in a crystal occupy quantum states with an infinite lifetime 
and band dispersion $\epsilon(\bk)$ set by the lattice potential.
Interactions with the other electrons and elementary excitations of the system
scatter the electrons, resulting in an altered dispersion relation $E(\bk)$ and a finite
lifetime. These many-body effects are encoded in the complex self-energy $\Sigma(\bk, E)
= \Sigma^{\prime}(\bk, E) + i\Sigma^{\prime\prime}(\bk, E)$. The imaginary
part $\Sigma^{\prime\prime}(\bk, E)$ determines the lifetime of the state and is
related to the scattering rate. 
The real part $\Sigma^\prime(\bk, E)$ shifts the electronic 
dispersion $E(\bk) = \epsilon(\bk) +  \Sigma^\prime(\bk, E)$. 
The tools available for studying energy and 
momentum resolved self-energy are limited.\cite{Carbotte} For example, bulk
transport and optical spectroscopies provide some access to  
$\bk$-integrated self-energies while ARPES accesses $\bk$-resolved 
information for only the occupied states.   
It is therefore important to develop a more extensive suite of versatile 
techniques, especially in the context of 
complex systems that remain poorly understood from a theoretical perspective.  

STM/STS accesses momentum space electronic structure by imaging real-space 
maps of the modulations in
differential conductance ($dI/dV$), which is proportional to the local
density of states (LDOS) of the sample. These modulations arise from the
interference of electrons scattered elastically by defects, and contain
information about the initial and final momenta that 
are accessible by a Fourier transform of the real space map. As the electrons
are dressed by interactions, the momentum space scattering intensity map
is often referred to as the quasiparticle interference (QPI) map. The dominant
intensities in a QPI map occur at scattering wave vectors linking constant
energy segments of the band dispersion. By tracking the energy dependence of
these peaks, the electronic dispersion $E(\bk)$ can be obtained. This technique has
been used to map coarse dispersions in many materials
\cite{Rutter,Roushan} 
and to examine scattering selection rules.\cite{Rutter,Roushan}
While the influence of many-body effects in  
FT-STS has been postulated since early reports,\cite{HengsbergerPRL1999} the direct influence on
dispersion and a quantitative account of self-energy effects were missing.
Here, we extend this technique to the examination of fine structure in the $E(\bk)$
due to many-body interactions. 


The two-dimensional Shockley surface state of the noble metal
silver Ag(111) was selected as an ideal system for demonstrating the
quantitative 
capabilities of FT-STS.  
The Ag(111) surface state is well characterized,   
and while it exhibits distinct many-body effects, it 
lacks the complicated interplay of interactions that appear in many complex
modern materials.
\cite{LiPRB1997, LiPRL1998,Paniago1995, EigurenPRL2002,
Buergi2000,Buergi2000_2,BuergiThesis,PaniagoiSurfaceScience1995,KliewerScience2000,Pennec,EELS} 
The dispersion of the surface state is free-electron-like over a wide energy
range given by $\epsilon(\bk) =
\hbar^2 k^2/2m^* - \mu$, where $m^*$ is the effective mass and $\mu$ is the 
chemical potential.  This parabolic dispersion is modified by the
electron-electron (e-e) and
electron-phonon (e-ph) interactions.  As both are accurately 
described by conventional theory,\cite{Buergi2000,BuergiThesis} a
straightforward comparison with theory can be made, requiring few  
parameters. The e-e interaction decreases the electron lifetime for energies
away from the Fermi level. The e-ph interaction introduces an additional
scattering channel 
for energies above the typical phonon energy scale (the Debye frequency
$\hbar\Omega_D$), 
decreasing the lifetime and modifying the bare dispersion near the 
Fermi energy $E_F$.  
The latter is a subtle effect in the case of Ag(111), that had not yet been 
observed due to the high resolution required in both energy and momentum.  

\begin{figure} 
\begin{center}
\includegraphics[width=0.8\textwidth]{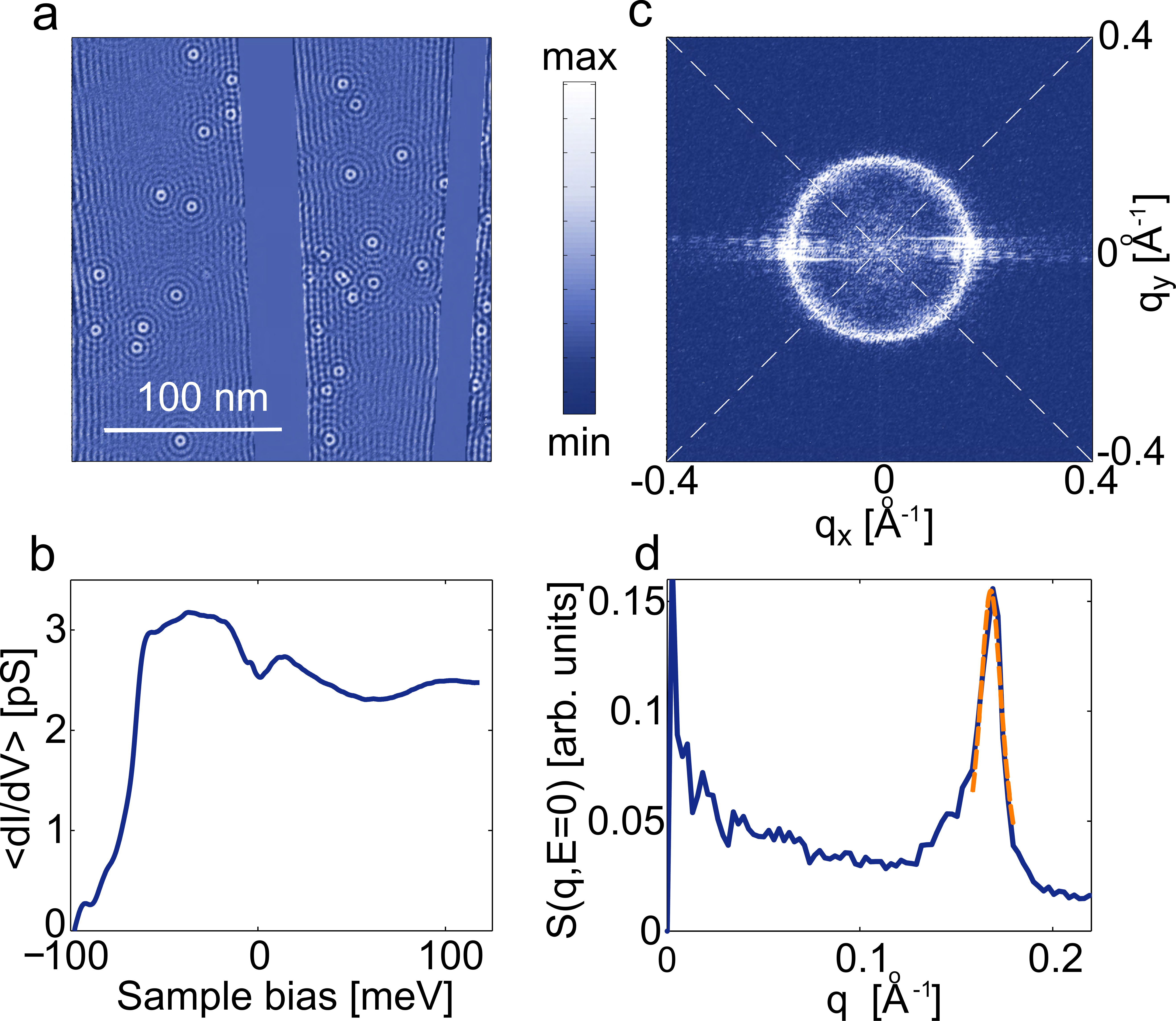}
\end{center}
\caption{\label{Fig:1} 
{\bf A summary of FT-STS of the surface state of Ag(111).}
(a) Conductance map (dI/dV) of a $239\times 239$ nm$^{2}$ area at $E=eV=0$ meV 
(tip height set at $V = 100$ meV, $I = 200$ pA). 
LDOS modulation due to scattering at step edges and CO adsorbates are
visible. The areas around the step edges were removed as discussed in the text.
Furthermore, a tip change induced stripe was corrected by a line-by-line
subtraction of the average line value that excludes CO impurities.  (b) Average
$dI/dV$ spectrum from a defect free area with a total size of
100 nm$^2$. The particle-hole symmetric steps at $E_F$ likely originate from an
inelastic co-tunneling pathway via phonon modes polarized perpendicular to the
surface. (c) Absolute value of the Fourier transform (power spectrum) of the
$dI/dV$ map ($E = 0$, panel a) showing a ring with radius $\bq=2\bk_F$ where $\bq$ is 
the scattering vector. The increased intensity along the $q_x$ direction 
originates from the step edge contributions.
(d) The QPI line profile $S(\bq , E=0)$. This is obtained by integrating (c) within
the range above and below the dashed lines in order to isolate contributions
from the CO adsorbates. The scattering peak is slightly asymmetric with an
enhanced intensity at low $\bq$, which is more pronounced at higher energies. 
Scattering peak positions and heights were obtained by fitting the line
profiles as shown in (a) within a range of $\pm0.01$ \AA$^{-1}$ around the
peak position (dashed line).
}
\end{figure}

STM/STS measurement of the Ag(111) surface yields real-space 
conductance maps (see Fig. \ref{Fig:1}a for the map at $E=E_F$) with 
circular LDOS modulations arising from scattering from point-like CO 
adsorbates, and vertical modulations produced by step-edge reflections. 
The small terraces on the surface produce subtle 
confinement effects not representative of the 
pristine surface state.  In order to access intrinsic surface properties, 
we removed their contribution by setting $dI/dV$ in this region to the average
value over the entire image.  The ability to isolate regions of interest in
this way is unique to STM/STS, as probes such as ARPES would average over these
domains.  Fig. 1b shows a typical $dI/dV$ spectra, averaged over a defect free
region.  The momentum space QPI intensity map $S(\bq,E_F)$ (Fig. \ref{Fig:1}c)
exhibits a ring of radius $\bq(E_F) = 2\bk(E_F)$, as expected for a
free-electron-like dispersion where back scattering is
dominant.\cite{Capriotti2003} A line profile $S(|\bq|,E_F)$ of the QPI map is
shown in Fig. \ref{Fig:1}d, where we have performed an angular average of 
$S(\bq,E_F)$ in the regions above and below the dashed lines. This restriction
isolates the contributions of the point-like CO scatterers.  
The momentum space resolution $\Delta q \sim0.0026$ $\AA^{-1}$ 
is set by the dimension of the map (
$239\times239$ nm$^2$) while the energy resolution $\Delta E = 4k_BT
= 1.5$ meV is limited by thermal broadening of the tip and the sample. 

\begin{figure}
\begin{center} 
\includegraphics[width=0.8\textwidth]{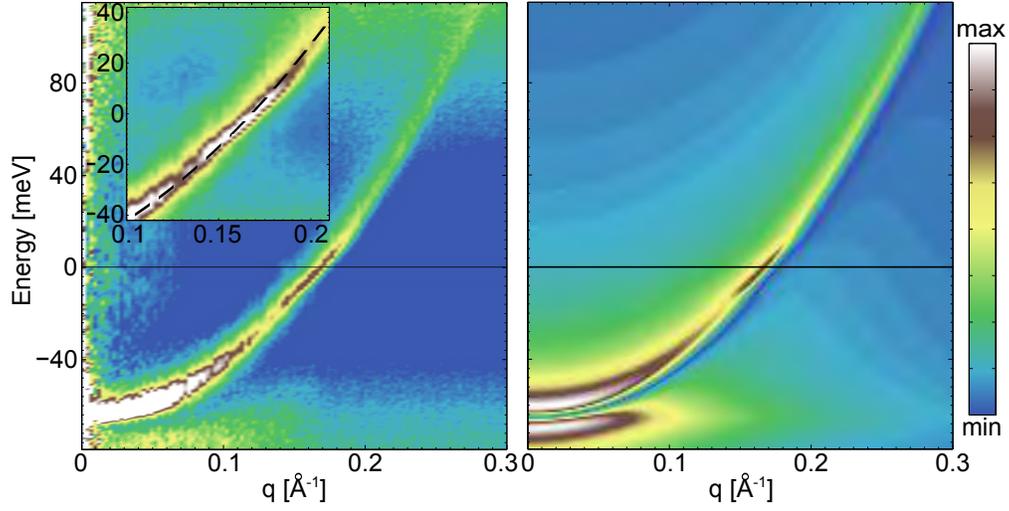}
\end{center}
\caption{\label{Fig:2} 
{\bf Measured and calculated dispersion of the QPI intensity.} 
(a) The measured dispersion of $S(\bq, E)$, obtained by plotting the line
profiles as shown in \ref{Fig:1}d for all bias voltages. 
Overall the dispersion is parabolic with $\mu=65 \pm 1$ meV and $m^*/m_e=0.41\pm0.02$,  
obtained by fitting the peak position excluding the energy range $[-20, 20]$ meV.
The intensity of the scattering peak generally decreases with increasing energy
but has a non-monotonic increase near $E_F$. We observe an additional
scattering intensity below the onset of the surface state ($E < -70$ meV).
The inset reveals a subtle renormalization of the dispersion within $E_F\pm14$
meV. (b) Theoretical calculation of the QPI intensity with $\mu=65$ meV,  $m^*/m_e=0.41$. 
The model includes electron-electron (Fermi-liquid theory) and
electron-phonon (Debye model, $\hbar\Omega_D$=14 meV, $\lambda=0.13$) interactions
and assumes the CO adsorbates scatter in the unitary limit.
}
\end{figure}

We now examine the detailed electronic structure of the Ag(111) surface state by 
considering the full energy dependence of  
the angle-averaged profile 
$S(|\bq|,E)$, as shown in Fig. \ref{Fig:2}a.  A parabolic band
dispersion is evident over a wide energy range while the QPI signal  
intensity exhibits a monotonic decrease  
from the onset of the surface state to higher energy. From the data   
we extract $\mu = 65 \pm 1$
meV and $m^∗/m_e = 0.41 \pm 0.02$, consistent with previous 
measurements.\cite{LiPRB1997,Paniago1995,Buergi2000,BuergiThesis}
These values define the bare dispersion in the absence of e-ph coupling. 
Deviations from this dispersion, as well as an enhanced QPI signal intensity, 
are evident in the vicinity of
$E_F$, shown more clearly in the inset of Fig. \ref{Fig:2}a.  
A similar increase in QPI signal intensity was observed near $E_F$ in one of
the first FT-STS reports on the Be(0001) surface state\cite{SprungerScience1997} 
and the e-ph interaction was later proposed as a possible origin.\cite{HengsbergerPRL1999}

To identify the source of these deviations and to assess the QPI signal 
we modeled the system using
the T-matrix formalism, considering scattering from a single
CO impurity in the unitary limit (see methods).  
The e-e interaction was handled within Fermi liquid theory while the 
e-ph interaction was treated within standard Migdal theory\cite{Migdal} 
with the phonons described by the Debye model (Debye frequency $\hbar\Omega_D
= 14$ meV, dimensionless e-ph coupling strength $\lambda = 0.13$).  
Within these approximations the self-energy $\Sigma(E)$ is a function 
of energy only. The
resulting simulated QPI intensity is shown in Fig \ref{Fig:2}b. The model 
closely reproduces both the coarse and fine details of the data.
The overall monotonic decrease in QPI intensity is linked 
to the group velocity of the bare electronic dispersion and is not related 
to many-body effects. However, the increase in the intensity near $E_F$ 
and the deviations from the parabolic band dispersion arise from 
the e-ph interaction. 

\begin{figure} 
\begin{center}
\includegraphics[width=0.5\textwidth]{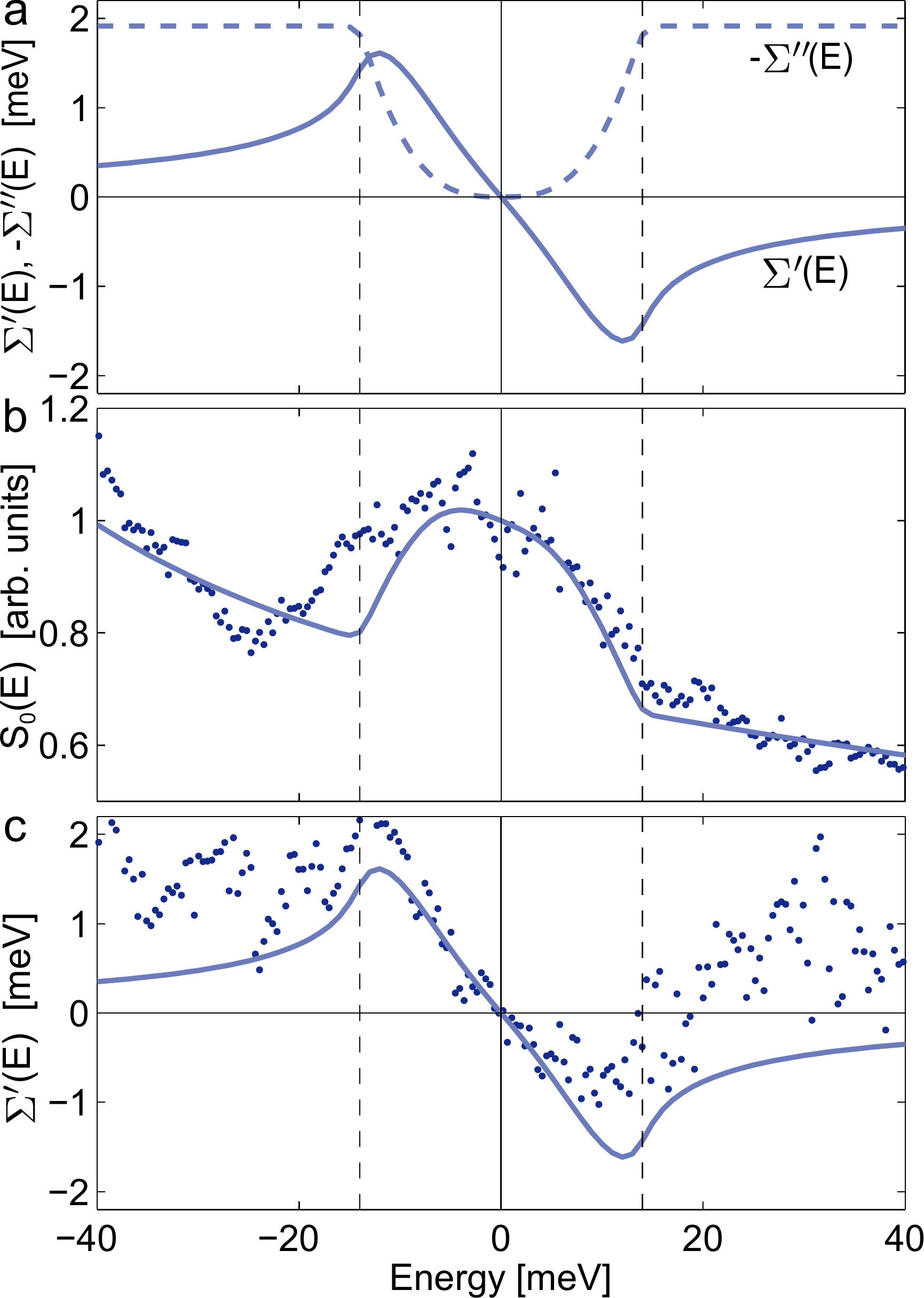}
\end{center}
\caption{\label{Fig:3} 
{\bf Quantitative extraction of the real and imaginary part of the self-energy.} 
(a) Calculated real ($\Sigma'$, solid line) and imaginary ($\Sigma''$, dashed line)
parts of the self-energy for the parameters used in Fig. \ref{Fig:2}b. 
(b) Scattering peak height $S_0(\bq,E)$ as a function of energy determined
experimentally (dots) and
theoretically (solid line). The theory curve has been normalized to the 
value at $E = 0$ while the experimental data had been normalized to the 
average value over the window $[-5,5]$ meV. 
The overall decrease of the scattering peak height with increasing energy is
related to the energy dependence of the group velocity. 
The increased intensity at $E_F\pm14$ meV is caused by a dip in $-\Sigma''$ 
due to the e-ph interaction. 
This relationship opens up a new way to experimentally obtain $\Sigma''$.
(c) Real part of the self energy determined from the difference between the
scattering peak position $E(\bk)$ and a parabolic fit $\epsilon(\bq)$.  The solid
line corresponds to the theoretically determined $\Sigma'$ as presented in (a).
Calculated and measured self-energy real part agree well, demonstrating that
FT-STS is an alternative tool to obtain $\Sigma'(\omega)$.  The dashed lines at $\pm$ 14
meV in (a) to (c) indicate the position of the Debye energy $\hbar\Omega_D$.  }
\end{figure}

We now perform a quantitative analysis of the self-energy. For reference, Fig. 
\ref{Fig:3}a shows the e-ph self-energy for the same values of 
$\hbar\Omega_D$ and $\lambda$ used in Fig. \ref{Fig:2}b. 
To extract $\Sigma(E)$, a Lorentzian was fit to the data in the vicinity of the
peak to obtain both the QPI peak position and height. (See
methods. An example fit is shown as the dashed line in Fig. \ref{Fig:1}d.) A plot of the
QPI peak height reflects the behavior of $\Sigma^{\prime\prime}(E)$ as shown 
in Fig. \ref{Fig:3}b, where we
compare the data with the model. 
Here we have normalized both sets of data (as described in the figure caption) in
order to eliminate the role of the tunneling matrix element in setting the scale of
the experimental data. 
There is good agreement between the model and experiment
apart from a slight deviation $\sim 20$ meV, which we attribute to a set point effect
below $\bq_F = 2\bk_F$.\cite{Buergi2000} 
The decrease in $-\Sigma^{\prime\prime}(E)$ within $\hbar\Omega_D = 14$ meV of $E_F$ 
produces the non-monotonic variation in peak height 
superimposed over the group velocity dependence.
This is due to the  
closing of the phonon scattering channel at energies below the 
characteristic phonon frequency, resulting in longer-lived quasiparticles near $E_F$.   
We note that the value $\hbar\Omega_D$ 
required to reproduce the data is close to the value for the top 
of the bulk acoustic branches.\cite{EELS}
The real part $\Sigma^\prime(E)$ can be estimated from the data by taking the
difference between
the measured peak position and the parabolic dispersion. The result is shown in
Fig. \ref{Fig:3}c, where peaks in $\Sigma^\prime(E)$ occur in the data at the same energy
scale reflected in Fig. \ref{Fig:3}b. The dimensionless strength of the e-ph 
coupling $\lambda$ can be estimated from 
$d\Sigma^\prime/dE|_{E=E_F}$.\cite{Migdal} We obtain $\lambda = 0.13 \pm 0.02$, 
consistent with previous estimates.\cite{EigurenPRL2002,Buergi2000}

Our FT-STS results provide a stunning visualization of the
subtle modifications in 
dispersion and scattering intensity arising from many-body interactions in a
simple system. This method provides high resolution in both momentum and
energy that is competitive with state-of-the-art ARPES.  Moreover, FT-STS
accesses both occupied and unoccupied states opening up the possibility of
examining particle-hole asymmetric systems.  These aspects give access to many-body
features not previously observed in such a direct way. With enhanced
stability and lower temperatures, further advancements in the application of
FT-STS to quantify many-body interactions in more complex systems can undoubtedly  
be expected.  However, perhaps the most compelling advantage of
FT-STS is the prospect of exploiting STM's unique spatial sensitivity to 
explore variations in the many-body interactions in
nanoscale regions and intrinsically inhomogeneous materials.

\begin{center}
{\bf Methods} 
\end{center}
{\bf Experiment:}
Measurements were performed in a Createc STM in ultra high vacuum at a temperature of 4.2 K with a
tungsten tip formed by direct contact with the Ag crystal.  The Ag(111) surface
was cleaned by three cycles of Ar sputtering each followed by thermal annealing
to 500 $^\circ$C.  The $I(V)$ map measured over 80 hours consists of
$380\times380$ spectra taken on a $239\times239$ nm$^2$ area. 
Each $I(V)$ spectrum consists of 512 data points and was Gaussian smoothed 
maintaining thermally limited energy resolution of $\Delta E=1.5$ meV. 
$dI/dV$ spectra were acquired by numerical differentiation of the $I$-$V$ sweep.
Atomic resolution scans on the same area lead to a spatial resolution of
(0.58$\pm$0.01) nm. 
The peak position $S_O$ of each line
profile $S(\bq,E)$ was determined by fitting a Lorentzian 
within a range of $\pm0.01$ \AA$^{-1}$ 
peak position expected from $\epsilon(\bk)$.  The renormalization of the bare
dispersion has been observed in four different data sets. 

{\bf Theory:}
Calculations were performed using the $T$-matrix formalism for scattering from
a single impurity.\cite{Capriotti2003} The QPI intensity is given by the
impurity-induced LDOS modulations 
\begin{equation} \delta\rho({\bf q},\omega) =
-\frac{i}{2\pi}\sum_{\bf k}{\mathrm{Im}} G({\bf k},\omega)T(\omega)G({\bf k} +
{\bf q},\omega).  
\end{equation} 
Here $T = -V_0\sin(\delta)\exp(i\delta)$ is
the $T$-matrix, where $\delta = \pi/2$ is the phase shift (unitary limit),
$V_0$ is the scattering potential.   The Green's function is given by $G({\bf
k},\omega) = [\omega - \xi({\bf k}) - \Sigma(\omega)]^{-1}$ where $\xi =
-\hbar^2 k^2/2m^* - \mu$ ($m^* = 0.41m_e,\mu = 65$ meV) is the dispersion of
the surface state and $\Sigma(\omega) = -i\eta - i\gamma\omega^2/2 +
\Sigma_{e-ph}(\omega)$ is the self-energy.  Here $\eta = 2.5$ meV is the lifetime
broadening due to scattering from the terraces,  $\gamma = 62.7$ meV is the e-e
interactions,\cite{BuergiThesis} and $\Sigma_{e-ph}(\omega)$ is 
obtained from a Debye model with $\hbar\Omega_D = 14$ meV 
and coupling strength $\lambda = 0.13$.\cite{BuergiThesis}     

\linespread{1}
\newpage

\newpage

\newpage

\begin{center}
{\bf Acknowledgements}
\end{center}
$^*$ These authors contributed equally to this work. 
The authors thank E. van Heumen for useful discussions.  
The authors gratefully acknowledge support by NSERC, CFI, CIFAR, the University of 
British Columbia, and the Canada Research Chairs program (S.B.). 
S.G. acknowledges support from The Woods fund.

\begin{addendum}
\item [Competing interests statement] The authors declare that they have no competing financial interests. 
\end{addendum}


\newpage
\begin{thebibliography}{99}
\bibitem{Migdal}
A. Migdal, 
``Interaction between electrons and lattice vibrations in a normal metal".
Sov. Phys. JETP {\bf 7}, 996 (1958).
\bibitem{Bostwick2006}
A. Bostwick, T. Ohta, T. Seyller, K. Horn, and E. Rotenberg, 
``Quasiparticle dynamics in graphene". 
Nature Physics {\bf 3}, 36 (2006).
\bibitem{Bostwick2010}
A. Bostwick, F. Speck, T. Seyller, K. Horn, M. Polini, R. Asgari, 
A. H. MacDonald, and E. Rotenberg, 
``Observation of Plasmarons in Quasi-Freestanding Doped Graphene". 
Science {\bf 328}, 999 (2010).
\bibitem{ZhuPRL2012}
Xuetao Zhu, L. Santos, C. Howard, R. Sankar, F. C. Chou, C. Chamon, 
and M. El-Batanouny, 
``Electron-Phonon Coupling on the Surface of the Topological Insulator 
Bi$_2$Se$_3$ Determined from Surface-Phonon Dispersion Measurements". 
Phys. Rev. Lett. {\bf 108}, 185501 (2012).
\bibitem{McMillan}
W. L. McMillan and J. M. Rowell, 
``Lead Phonon Spectrum Calculated from Superconducting Density of States". 
Phys. Rev. Lett. {\bf 14}, 108 (1965).
\bibitem{Lanzara}
A. Lanzara, P. Bogdanov, X. Zhou, S. Kellar, D. Feng,
E. Lu, T. Yoshida, H. Eisaki, A. Fujimori, K. Kishio, {\it et al.},
``Evidence for ubiquitous strong electron-phonon coupling in high-temperature superconductors". 
Nature {\bf 412}, 510 (2001).
\bibitem{Dahm}
T. Dahm, V. Hinkov, S. Borisenko, A. Kordyuk,
V. Zabolotnyy, J. Fink, B. B{\"u}chner, D. Scalapino,
W. Hanke, and B. Keimer, 
``Strength of the spin-fluctuation-mediated pairing interaction in a high-temperature superconductor". 
Nature Physics {\bf 5}, 217 (2009).
\bibitem{Byczuk}
K. Byczuk, M. Kollar, K. Held, Y.-F. Yang, I. Nekrasov, T. Pruschke, 
and D. Vollhardt, 
``Kinks in the dispersion of strongly correlated electrons". 
Nature Physics {\bf 3}, 168 (2007).
\bibitem{Graf}
J. Graf, G.-H. Gweon, K. McElroy, S. Zhou, C. Jozwiak, E. Rotenberg, 
A. Bill, T. Sasagawa, H. Eisaki, S. Uchida, {\it et al.}, 
``Universal High Energy Anomaly in the Angle-Resolved 
Photoemission Spectra of High Temperature Superconductors: 
Possible Evidence of Spinon and Holon Branches"
Phys. Rev. Lett. {\bf 98}, 067004 (2007).
\bibitem{FiskNature1986}
Z. Fisk, H. Ott, T. Rice, and J. Smith, ``Heavy electron metals". 
Nature {\bf 320}, 124 (1986).
\bibitem{SprungerScience1997}
P. Sprunger, L. Petersen, E. Plummer, E. Lægsgaard, and F. Besenbacher, 
``Giant Friedel Oscillations on the Beryllium(0001) Surface". 
Science {\bf 275}, 1764 (1997).
\bibitem{Petersen}
L. Petersen, P. T. Sprunger, P. Hofmann, E. Lægsgaard, B. G. Briner, 
M. Doering, H.-P. Rust, A. M. Bradshaw, F. Besenbacher, and E. W. Plummer,
``Direct imaging of the two-dimensional Fermi contour: Fourier-transform STM". 
Phys. Rev. B {\bf 57}, R6858 (1998).  
\bibitem{McElroyNature2003}
K. McElroy, R. Simmonds, J. Hoffman, D.-H. Lee, J. Orenstein, H. Eisaki, 
S. Uchida, and J. Davis, 
``Relating atomic-scale electronic phenomena to wave-like quasiparticle states in superconducting 
Bi$_2$Sr$_2$CaCu$_2$O$_{8+\delta}$".
Nature {\bf 422}, 592
(2003)
\bibitem{HoffmanScience2002}
J. E. Hoffman, K. McElroy, D.-H. Lee, K. M. Lang, H. Eisaki, S. Uchida, 
and J. C. Davis, 
``Imaging Quasiparticle Interference in Bi$_2$Sr$_2$CaCu$_2$O$_{8+x}$".
Science {\bf 297}, 1148 (2002).
\bibitem{Carbotte}
J. P. Carbotte, T. Timusk, and J. Hwang, 
``Bosons in high-temperature superconductors: an experimental survey".  
Rep. Prog. Phys. {\bf 74}, 066501 (2011).
\bibitem{Rutter}
G. M. Rutter, J. N. Crain, N. P. Guisinger, T. Li, P. N. First, 
and J. A. Stroscio, 
``Scattering and interface in Epitaxial Graphene". 
Science {\bf 317}, 219 (2007),
\bibitem{Roushan}
P. Roushan, J. Seo, C. V. Parker, Y. Hor, D. Hsieh, D. Qian, 
A. Richardella, M. Z. Hasan, R. Cava, and A. Yazdani, 
``Topological surface states protected from backscattering by chiral spin texture". 
Nature {\bf 460}, 1106 (2009).
\bibitem{LiPRB1997} 
J. Li, W.-D. Schneider, and R. Berndt,
``Local density of states from spectroscopic scanning-tunneling-microscope images: Ag(111)". 
Phys. Rev. B {\bf 56}, 7656 (1997).
\bibitem{LiPRL1998}
J. Li, W.-D. Schneider, R. Berndt, O. R. Bryant, and 
S. Crampin, 
``Surface-State Lifetime Measured by Scanning Tunneling Spectroscopy"
Phys. Rev. Lett. {\bf 81}, 4464 (1998).
\bibitem{Paniago1995} 
R. Paniago, R. Matzdorf, G. Meister, and A. Goldmann,
``Temperature dependence of Shockley-type surface energy bands on Cu(111), Ag(111) and Au(111)".
Surface Science {\bf 336}, 113 (1995).
\bibitem{EigurenPRL2002}
A. Eiguren, B. Hellsing, F. Reinert, G. Nicolay, E. Chulkov,
V. Silkin, S. H{\"u}fner, and P. Echenique, 
``Role of Bulk and Surface Phonons in the Decay of Metal Surface States". 
Phys. Rev. Lett. {\bf 88}, 066805 (2002).
\bibitem{Buergi2000_2}
L. B{\"u}rgi, L. Petersen, H. Brune, and K. Kern, 
``Noble metal surface states: deviations from parabolic dispersion". 
Surface Science {\bf 447}, L157 (2000).
\bibitem{Buergi2000}
L. B{\"u}rgi, H. Brune, O. Jeandupeux, and K. Kern, 
``Quantum coherence and lifetimes of surface-state electrons". 
Journal
of Electron Spectroscopy and Related Phenomena {\bf 109}, 33
(2000).
\bibitem{BuergiThesis}
L. B{\" u}gri, ``Scanning tunneling microscopy as local probe of electron 
density, dynamics and transport at metal surfaces". Ph.D. Thesis (1999).
\bibitem{PaniagoiSurfaceScience1995}
R. Paniago, R. Matzdorf, G. Meister, and A. Goldmann,
``High-resolution photoemission study of the surface states near $\bar{\Gamma}$ on Cu(111) and Ag(111)". 
Surface Science {\bf 331}, 1233 (1995).
\bibitem{KliewerScience2000}
J. Kliewer, R. Berndt, E. Chulkov, V. Silkin, P. Echenique,
and S. Crampin, 
``Dimensionality Effects in the Lifetime of Surface States". 
Science {\bf 288}, 1399 (2000).
\bibitem{Pennec}
Y. Pennec, W. Auw{\"a}rter, A. Schiffrin, A. Weber-Bargioni, A. Riemann, 
and J. Barth, ``Supramolecular gratings for tuneable confinement of electrons on metal surfaces". 
Nature Nanotechnology {\bf 2}, 99 (2007).
\bibitem{EELS}
L. Chen, L. L. Kesmodel, and J.-S. Kim, ``EELS studies of surface phonons on Ag(111)". 
Surface Science {\bf 350}, 215 (1996).  
\bibitem{HengsbergerPRL1999}
M. Hengsberger, D. Purdie, P. Segovia, M. Garnier, and Y. Baer, 
``Photoemission Study of a Strongly Coupled Electron-Phonon System". 
Phys. Rev. Lett. {\bf 83}, 592 (1999).
\bibitem{Capriotti2003}
L. Capriotti, D.J. Scalapino, and R.D. Sedgewick, 
``Wave-vector power spectrum of the local tunnelling density of states: ripples in a d-wave sea". 
Phys. Rev. B {\bf 68}, 014508 (2003).
\end{thebibliography}
\end{document}